# Magnetization Characteristic of Ferromagnetic Thin Strip by Measuring Anisotropic Magnetoresistance and Ferromagnetic Resonance

Ziqian Wang, Guolin Yu, Xinzhi Liu, Bo Zhang, Xiaoshuang Chen and Wei Lu

*National Laboratory for Infrared Physics, Shanghai Institute of Technical Physics, Chinese Academy of Sciences, Shanghai, 200083, China*

**Abstract**

The magnetization characteristic in a permalloy thin strip is investigated by electrically measuring the anisotropic magnetoresistance and ferromagnetic resonance in in-plane and out-of-plane configurations. Our results indicate that the magnetization vector can rotate in the film plane as well as out of the film plane by changing the intensity of external magnetic field of certain direction. The magnetization characteristic can be explained by considering demagnetization and magnetic anisotropy. Our method can be used to obtain the demagnetization factor, saturated magnetic moment and the magnetic anisotropy.

## 1. Introduction

Anisotropic magnetoresistance (AMR) effect, which is resulted from the anisotropy of spin-orbit interaction in ferromagnetic materials [1,2], was first discovered by Thomson in 1857 [3]. The effect bears an essential role for both scientific perspectives and technological applications [4-8]. AMR is manifested in the dependence of the resistivity on the angle between current and magnetization direction [1,2], and is given by

$$R = R_0 + R_A - R_A \sin^2 \theta. \tag{1}$$

Here $R_0$ represents the resistance while the magnetization **M** is perpendicular to the induced current, $R_A$ is the decrement resistance, and θ is the angle of magnetization **M** with respect to current. In ferromagnetic devices of certain structures, **M** is parallel to the effective magnetic field $\mathbf{H_{eff}}$, including external field $\mathbf{H_{ex}}$, anisotropy built-in field $\mathbf{H_a}$ and demagnetization field $\mathbf{H_d}$. Researchers can obtain these parameters through finding θ.

The purpose of this article is to obtain magnetization characteristic of a ferromagnetic thin strip through AMR measurement. Ferromagnetic resonance (FMR) detection is considered as an ancillary method to determine the magnetization based on fitting the measured FMR dispersion curves via Kittel's theory [9].

2. Material and Methods

The present work is performed on a $Ni_{80}Fe_{20}$ (Permalloy, Py) thin film deposited on a 5×6 mm$^2$ GaAs single crystal substrate. This polycrystalline structured Py film is patterned to a stripe shape by photolithography and lift off techniques. The dimensions of our sample is: length=2400μm, width=200μm and thickness=50nm.

The Py strip is fixed on a rotatable holder by an adjustable wedge. The external magnetic field, represented as $\vec{H}_{ex} = (H_x,\ H_y,\ H_z)$, encloses an intersection angle α with the long-axis of the strip, and β represents the dip of the wedge. Accurate α and β are recorded by a readout on the holder and a goniometer. AMR is measured by detecting the resistance between two electrodes at each side of the strip's length, as illustrated in Fig. 1(a). In FMR measurement, modulated microwaves are propagating to the holder normally through a rectangular waveguide of X band, while the modulation frequency is 5.37 kHz. The FMR signals are also electrically measured in field-swept mode by using a lock-in amplifier connecting those two electrodes via gold bonding wires and coaxial cables.

The coordinate system we select in this article is demonstrated in Fig. 1(b). The long axis of the thin strip is set as z-axis, the direction perpendicular to the strip's plane is defined as y-axis, and the strip lies in the xz-plane. $\Phi_H$, $\Phi_M$ are recorded as the misalignments of $\mathbf{H_{ex}}$ and $\mathbf{M}$ with respect to xz-plane. The in-plane components of $\mathbf{H_{ex}}$ and $\mathbf{M}$ encloses the angles $\Theta_H$, $\Theta_M$ with z-axis, respectively. Hence $\Theta_H$, $\Phi_H$, $\alpha$ and $\beta$ follows the relations as $\Theta_H = \tan^{-1}(\tan\alpha\cos\beta)$ and $\Phi_H = \sin^{-1}(\sin\alpha\sin\beta)$.

In section 3.1, AMR and FMR measurements in weak $\mathbf{H_{ex}}$ are illustrated, and both of $\mathbf{H_a}$ and $\mathbf{H_d}$ are taken into account. We will show how $\mathbf{M}$ rotates from parallel to $\mathbf{H_a}$ via only changing the magnitude of $\mathbf{H_{ex}}$ generated by an electromagnet at room temperature. The detailed process of obtaining the demagnetization factors and magnetization through AMR and FMR experiments is also introduced in this subsection. For further investigating, AMR in out-of-plane configuration in stronger $\mathbf{H_{ex}}$ is discussed in section 3.2. In this subsection, the Py thin strip is placed in $\mathbf{H_{ex}}$ produced by a cryomagnet, which is carried out at liquid helium temperature.

## 3. Results and Discussion
## 3.1. Weak External Field Condition

It is noted as "weak field condition" when the magnitude of $\mathbf{H_{ex}}$ is smaller than 2000 Oe. The measured AMR and FMR data of this Py thin strip in in-plane and out-of-plane configurations are shown in Fig. 2. Meanwhile, the demagnetization coefficient and **M** are obtained via fitting these experimental results by selecting appropriate models. Fig. 2(a) shows the measured sheet resistance R versus $\mathbf{H_{ex}}$ at different $\Theta_H$ in in-plane magnetized configuration. It is illustrated that R reaches its maximum at $\mathbf{H_{ex}}$=0, hence **M** is parallel, or anti-parallel, to z-axis as a higher resistance state without external magnetic field. The increment of R achieved by decreasing the magnitude of $\mathbf{H_{ex}}$ implies **M**'s rotation from parallel with $\mathbf{H_{ex}}$ to z-axis, as demonstrated in Eq. (1). The direction of **M** at zero-field state is caused by $\mathbf{H_a}$, the anisotropic built-in field. $\mathbf{H_a}$ is along with z-axis because of the lowest free energy for thin strip structure in this direction.

Here we assume $\mathbf{H_a}$ as a static filed, which is parallel to z-axis and is expressed as $\vec{H}_a = (0,\ 0,\ H_a)$. The effect of other important factor on the rotation of **M** is $\mathbf{H_d}$, the demagnetization field. $\mathbf{H_d}$ depends on the direction and magnitude of **M**, it is written by $\vec{H}_d = -(N_{xx}M_x,\ N_{yy}M_y,\ N_{zz}M_z)$, where $N_{xx}$, $N_{yy}$, and $N_{zz}$ are the demagnetization factors in x, y and z directions. The demagnetization factors satisfy the correlation of $N_{xx} + N_{yy} + N_{zz} = 1$ for *SI* and $N_{xx} + N_{yy} + N_{zz} = 4\pi$ for *CGS* [11]. Accordingly, the yielded relations between **M** and **H**$_{ex}$ are:

$$\frac{H_x - N_{xx}M_x}{M_x} = \frac{H_y - N_{yy}M_y}{M_y} = \frac{H_z + H_a}{M_z},$$

$$H_x = |\vec{H}_{ex}|\cos\Phi_H \sin\Theta_H, H_x = |\vec{H}_{ex}|\sin\Phi_H, H_z = |\vec{H}_{ex}|\cos\Phi_H \cos\Theta_H, \quad (2)$$

$$M_x = |\vec{M}|\cos\Phi_M \sin\Theta_M, M_y = |\vec{M}|\sin\Phi_M, M_z = |\vec{M}|\cos\Phi_M \cos\Theta_M$$

The demagnetization field along z-axis is neglected since $N_{zz}$ is much smaller than $N_{xx}$ and $N_{yy}$ in thin strip structure, as shown in Fig. 1(b). It is difficult to provide analytical solutions for Eqs. (2), however getting numerical solutions is not a hard task. We use the numerical method to fit the experiment data. In in-plane configuration under weak field condition, $\Phi_H=\Phi_M=0$ and $H_y=0$, and Eqs. (2) are transformed as,

$$\frac{|\vec{H}_{ex}|\sin\Theta_H - N_{xx}|\vec{M}|\sin\Theta_M}{|\vec{M}|\sin\Theta_M} = \frac{|\vec{H}_{ex}|\cos\Theta_H + H_a}{|\vec{M}|\cos\Theta_M} \quad . \quad (3)$$

Considering $\Phi_H=\Phi_M=0$, we have $\theta=\Theta_H$. Taking the numerical results of Eq. (3) into Eq. (1), **H**$_a$ and $N_{xx}M_x$ are obtained. $R_0$ and $R_A$ can be recorded directly through Fig. 1 as $R_0$=87.7 Ω and $R_a$=1.5Ω. Other fitted results are $H_a$=1.95Oe and $N_{xx}M_x$=5.1Oe.

The demagnetization factor $N_{yy}$ is significantly larger than $N_{xx}$ according to the thin strip structure. If **H**$_{ex}$ consists of y-component, **M** should be misaligned away to **H**$_{ex}$, which causes the surface magnetic charges in each side of xz-plane [10]. These surface magnetic charges generate a demagnetization field inside the sample with its direction opposite to the y-component of **H**$_{ex}$, and finally prohibit the misalignment of **M**. According to the shape of our sample, the demagnetization field generated by a very slight y-component of **M** can offset the y-component of **H**$_{ex}$. Thus, only the xz-component of **H**$_{ex}$ is worth to be considered, the motion of **M** according to changing **H**$_{ex}$ is almost in-plane, and we have,

$$\frac{|\vec{H}_{ex}|\cos\Phi_H \sin\Theta_H - N_{xx}|\vec{M}|\sin\Theta_M}{|\vec{M}|\sin\Theta_M} = \frac{|\vec{H}_{ex}|\cos\Phi_H \cos\Theta_H + H_a}{|\vec{M}|\cos\Theta_M}, \quad (4)$$

$\Theta_M \approx \theta$.

The larger out-of-plane component of **H**$_{ex}$ can be provided by increasing the dip angle β of the wedge. Comparating to the in-plane **H**$_{ex}$, larger out-of-plane **H**$_{ex}$ is needed for obtaining the same **H**$_{eff}$ and θ, as showing in Fig. 2(b).

The calculated magnitudes of **M** is obtained by fitting FMR experiment. The dispersion curves in different configurations are recorded in Fig. 2(c). For an in-plane **M** assisted with microwave in a magnetic field **H**, the resonant frequency $f_r$ for the *rf* signal is given by $2\pi f_r = \gamma\sqrt{[H+(N_{yy}-N_{zz})M][H+(N_{xx}-N_{zz})M]}$ [9]. In our sample, we have $H = H_{eff} = |\vec{H}_{ex}+\vec{H}_d+\vec{H}_a|$ and $\gamma = 181\mu_0$ GHz/T for Py, here $\mu_o$ is the permeability of vacuum. The **H**$_{ex}$'s range is from 1000Oe to 1800Oe, such amplitudes are much smaller than N$_{yy}$M**, and** consequently a very slight $\Phi_H$ can generate large enough **H**$_d$ to overcome the y-component of $H_{ex}$ unless the in-plane component of $H_{ex}$ is significantly smaller than its y-component. For the schematic of β= 0, 45°and 70° in Fig. 2(c) for $H_{ex}$ larger than 1000Oe, $N_{xx}M_x$ and $H_a$ are ignorable, and the simplified expression of dispersion curve is given by:

$$2\pi f_r = \gamma \sqrt{\left(H_{ex}\cos\Phi_H + N_{yy}M\right)H_{ex}\cos\Phi_H}. \tag{5}$$

In weak field condition, since $H_{ex}$'s y-component is offseted by $H_d$, higher $H_{ex}$ is needed in order to achieve high enough resonant $H_{eff}$ for larger β while the assisted microwave's frequency is fixed. Eq. (5) is used to fit the electrically detected FMR dispersion curves in Fig. 2(c) at different β. The fitted magnetization is M=10750Oe, and the demagnetization factor along with x-axis is calculated as $N_{xx}$=0.00047.

### 3.2. Strong External Field Condition

According to Eq. (2), larger $H_{ex}$ may provide y-component for $H_{eff}$, and the direction of **M** can be tilted away from xz-plane. In terms of $\Phi_M \neq 0$ for this situation, Eq. (1) would be revised as,

$$R = R_0 + R_A - R_A\left(\sin^2\Phi_M + \cos^2\Theta_M \cot^2\Theta_M \cos^2\Phi_M\right). \tag{6}$$

Here $\Phi_M$ and $\Theta_M$ are deduced from Eqs. (2). The magnetization characteristic of Py thin strip under stronger $H_{ex}$ is investigated in liquid helium temperature around 4.2K. Although $R_0$, $R_A$ and **M** vary at different temperatures, AMR feature of this Py thin strip in low temperature and $H_{ex}$ up to 5 T evolves as the prediction of Eqs. (2), seeing in Fig. 3.

The movement of **M** from z-axis to $H_{ex}$ is separated by two steps, the in-plane magnetization as investigated in the former paragraphs, and the out-of-plane magnetization while is $H_{ex}$ high enough. The movement of **M** can be illustrated by an approximate picture. $\Phi_M \approx 0$ in the first step, **M** rotates rapidly from $\Theta_M=0$ to $\Theta_M=\Theta_H$. The process is displayed in the embedded picture of Fig. 3. In the second step, larger xz-component of $H_{ex}$ only keeps $\Theta_M=\Theta_H$. However, $\Phi_M$ increases with stronger y-component of $H_{ex}$, as shown in Fig. 3.

Meanwhile, we should indicate that the two-step magnetic movement does not exist in every applied out-of-plane $H_{ex}$. Taking $\Phi_H=90°$ as an instance, the movement of **M** only includes out-of-plane step in yz-plane because $H_{ex}$ contributes no x-component field to the effective field.

## 4. Conclusions

In summary, we have demonstrated how the magnetization characteristic of a ferromagnetic thin strip changes in different external magnetic field based on the AMR and FMR measurements by considering demagnetization and magnetic anisotropy. It is shown that the magnetization vector can rotate in the film plane as well as out of the film plane by sweeping the intensity of external magnetic field, while the direction of external field is fixed. The out-of-plane AMR's low-temperature and high-field features are also well explained. Our method can be used to obtain the demagnetization factor, saturated magnetic moment and the magnetic anisotropy.

## Acknowledgement

The work is supported by the State Key Program for Basic Research of China (2013CB632705, 2011CB922004), the National Natural Science Foundation of China (10990104).

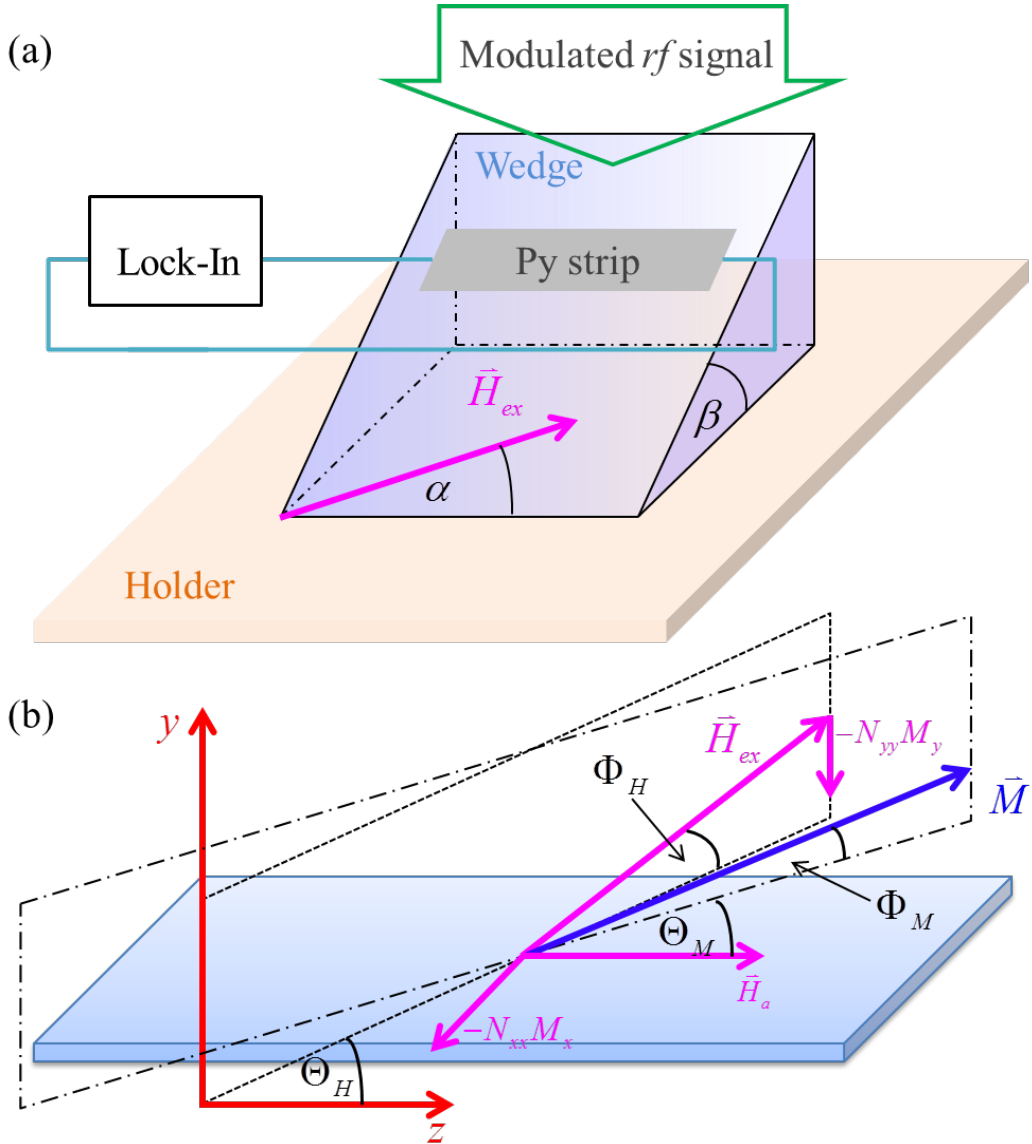

Figure 1. (a) The schematic of our experiment and (b) the coordinate system used in this article. The *rf* signal and Lock-In amplifier are only applied in FMR measurement. The demagnetization field along x and y direction are also notified as $-N_{xx}M_x$ and $-N_{yy}M_y$.

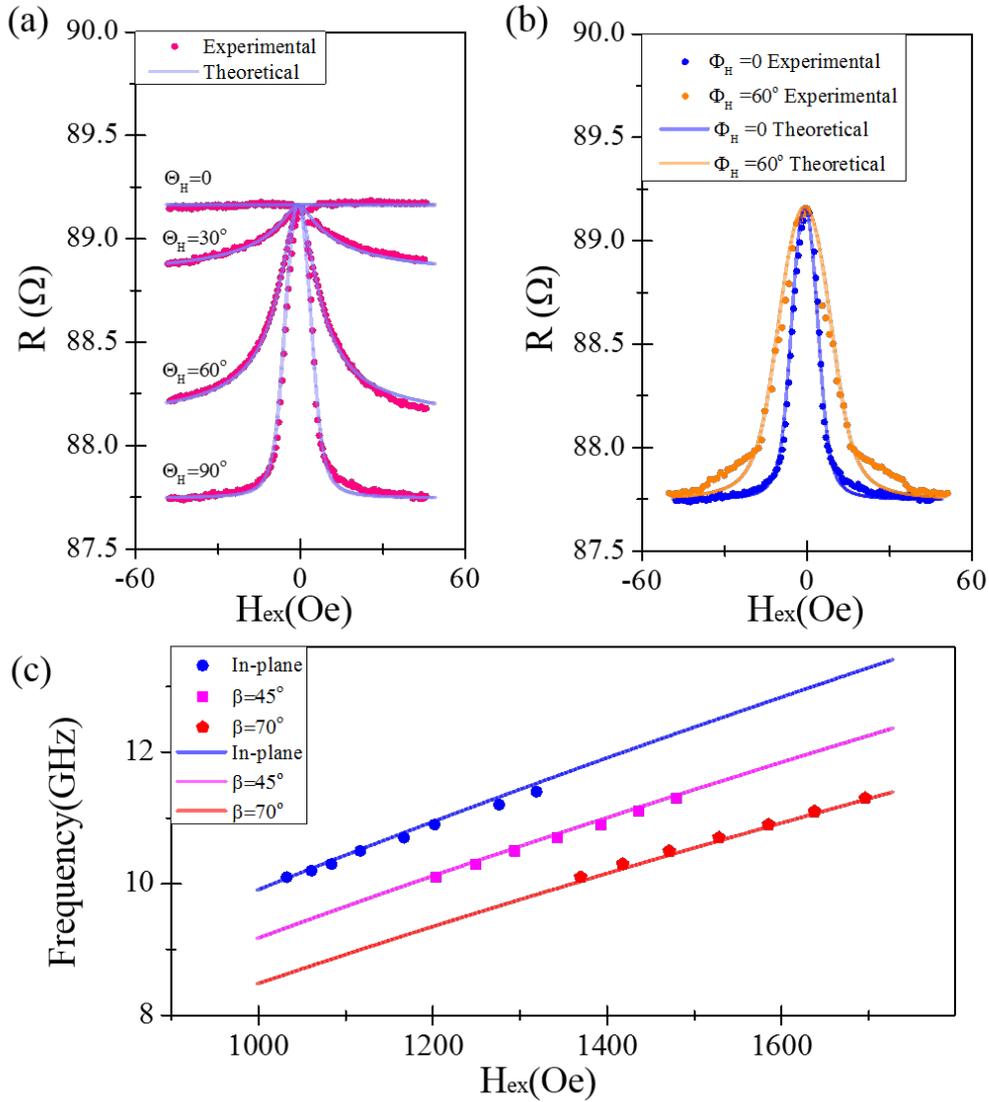

Figure 2. The experimental (colored dots) and fitted (solid lines) results for (a). in-plane magnetized AMR at different $\Theta_H$, (b). AMR of in-plane and out-of-plane magnetizations for $\Theta_H=90°$ at different $\Phi_H$, and (c). the resonant frequency of FMR with respect to external field for both in-plane and out-of-plane configurations at different β, here α is fixed at 45°.

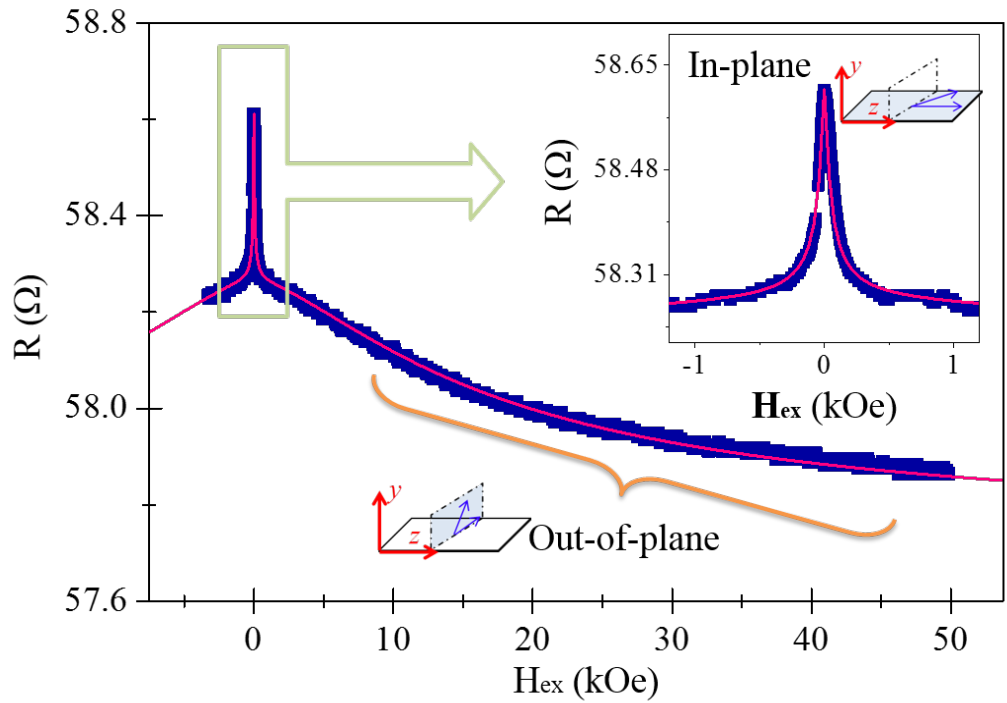

Figure 3. AMR measurement in stronger H$_{ex}$ at 4.2K. α and β are set as 45° and 30°, respectively. Inset: detailed sheet resistance of this thin strip at lower external field.